# Electronic and Magnetic Properties of Zigzag Boron-Nitride Nanoribbons with Even and Odd-line Stone-Wales (5-7 pair) Defects

SRKC Sharma Yamijala, [1] and Swapan K Pati [2, 3, *]

1. *Chemistry and Physics of Materials Unit, Jawaharlal Nehru Centre for Advanced Scientific Research, Jakkur P.O., Bangalore – 560064, India.*

2. *Theoretical Sciences Unit, Jawaharlal Nehru Centre for Advanced Scientific Research, Jakkur P.O., Bangalore – 560064, India.*

3. *New Chemistry Unit, Jawaharlal Nehru Centre for Advanced Scientific Research, Jakkur P.O., Bangalore – 560064, India.*

**Abstract:**

Spin-polarized first-principles calculations have been performed on zigzag Boron-Nitride Nanoribbons (z-BNNRs) with lines of alternating fused pentagon (P) and heptagon (H) rings (Pentagon-Heptagon-line-defect) at single edge as well as at both edges. The number of line (n) of the Pentagon-Heptagon-defect has been varied from 1 to 8 for 10-zBNNRs. Among the different spin-configurations which we have studied, we find that, the spin-configuration with ferromagnetic ordering at each edge and anti-ferromagnetic ordering across the edges is quite interesting. For this spin-configuration, we find that, if the introduced PH-line-defect is odd numbered, the systems behave as spin-polarized semi-conductors, but, for even numbered, all the systems show interesting anti-ferromagnetic half-metallic behavior. Robustness of these results has been cross checked by the variation of the line-defect position and also by the variation of the width [from ~ 1.1 nm (6-zBNNR) to ~ 3.3 nm (16-zBNNR)] of the ribbon. Density of States (DOS), projected-DOS and band-structure analysis have been accomplished to understand the reasons for these differences between even and odd-line-defects. The main reason for many of the observed changes was traced back to the change in edge nature of the BNNR, which indeed dictates the properties of the systems.

## I. INTRODUCTION

Nanoribbons, in particular graphene nanoribbons (GNRs), boron-nitride nanoribbons (BNNRs) and their hybrids, have attracted huge attention in recent years, because of their exciting applications in several fields like optics, electronics, opto-electronics, spintronics etc.[1-8] These applications of nanoribbons (NRs) mainly depend on their edges, and based on their edge nature, NRs are divided into two types viz., armchair and zigzag[5].

Between the zigzag (z) and armchair edged NRs, zigzag edged NRs are more interesting because of the presence of enhanced local-density of states near the Fermi-level, which in turn arise because of the localized electronic states at their edges[6]. Also, as proved by the previous studies on zGNRs[5, 8, 9] and zBNNRs[3, 10, 11], these edge-states can be tuned, using different methods viz., passivation, doping, substitution, and application of external electric-fields etc. to achieve tunable electronic and magnetic properties in these ribbons. Even, half-metallicity has been predicted in both zGNRs[7, 8, 12] and zBNNRs[1-3, 13] under different conditions.

It is important to mention that both GNRs and BNNRs have been successfully synthesized experimentally either by cutting the corresponding 2D structures viz., graphene[14] and BN sheets or by unzipping the corresponding nanotubes (NTs) viz., Carbon-NTs[15] and BN-NTs[16], respectively. But, achieving perfect edges experimentally is a difficult task and researchers have proved, experimentally, the presence of 5-7 edge reconstructions with fair abundance[17]. Recently, Huang et. al.[18] have shown the presence of 5-7-defect lines across the grain boundary in graphene. Very recently, Pan et. al.[19] also proved the presence of reconstructed 5-7 edges based on both the theory and experiment, in chiral GNRs. Apart from the 5-7 edge reconstructions, researchers have also observed the dynamics of the 5-7 defects inside the graphene[20, 21]. Also, Auwärter et. al.[22] have shown the two h-BN domains can co-exist on Ni (111) surface and they have demonstrated that linear defects define the boundary between these two domains, which also suggests the possibility of the presence of the 5-7 defects at these grain boundaries.

All the above experiments suggest that 5-7 defects are ubiquitous in graphene and its BN-analogues, and that the edges are more prone to these defects. So, it is required to know how the properties of the graphene or GNRs or BNNRs will change in the presence of these edge reconstructions. Theoretical studies have been performed on graphene[23, 24], GNRs[9] and also on BNNRs[10] to answer this question. The main result of all these studies is: "A 5-7 edge reconstruction on a nanoribbon with a zigzag-edge will bring stability to the system and it will destroy the magnetic property of the edge where the reconstruction has occurred". But, most of these studies have considered only one-line-PH-defect at one of the edges of a nanoribbon. Except Jia et. al., who proposed about two-line-PH-defect,[25] and a theoretical study by Sudipta et. al.,[9] who have considered both one-line and two-line-PH-defects at the edges of a GNR. In fact, there is no other study, in both GNRs and BNNRs, where the effect of the number of PH-line-defects in nanoribbons has been considered.

In this work, we have mainly investigated the structural, electronic and magnetic properties of zigzag boron-nitride nanoribbons (zBNNRs) with and without pentagon-heptagon (PH) line-defects. We have varied the PH-line-defect number from one to eight and we have shown that, it is possible to tune the properties of zBNNRs, from insulating – semiconducting – half-metallic, depending on the PH-line-defect number. We find that the behavior of the system is completely dependent on the nature of the line-defect (i.e. even or odd). Robustness of the observed properties has been verified by varying the width of the zBNNRs and also by varying the position of a PH-line-defect from one-edge to the other. The rest of the work is arranged as follows: In section II, we have presented the computational details. Section III describes the different kinds of systems which we have considered in the present work. Section IV is devoted to discuss the stability, electronic and magnetic properties of all these systems. Finally, in section IV, we summarize the main findings of the present work.

## II. COMPUTATIONAL DETAILS

We have performed the spin-polarized Density Functional Theory (DFT) calculations, for all our systems, using the ab-initio software package SIESTA[26, 27]. The generalized-gradient-approximation (GGA) with the Perdew-Burke-Ernzerhof (PBE)[28, 29] form is chosen for the exchange-correlation functional. Interaction between the ionic cores and the valence electrons is accounted by the norm conserving pseudo-potentials[30] in the fully non-local Kleinman-Bylander form[31]. The pseudo-potentials are constructed from 1, 3 and 5 valence electrons for the H, B and N atoms, respectively. To expand the wave-functions, numerical localized combination of atomic orbitals with double-$\zeta$ basis sets are used. To represent the charge density, a reasonable mesh-cut-off of 300 Ry is used for the grid integration. Conjugate-gradient (CG) method has been used to optimize the structures. We have also optimized the lattice parameters by allowing them to change.

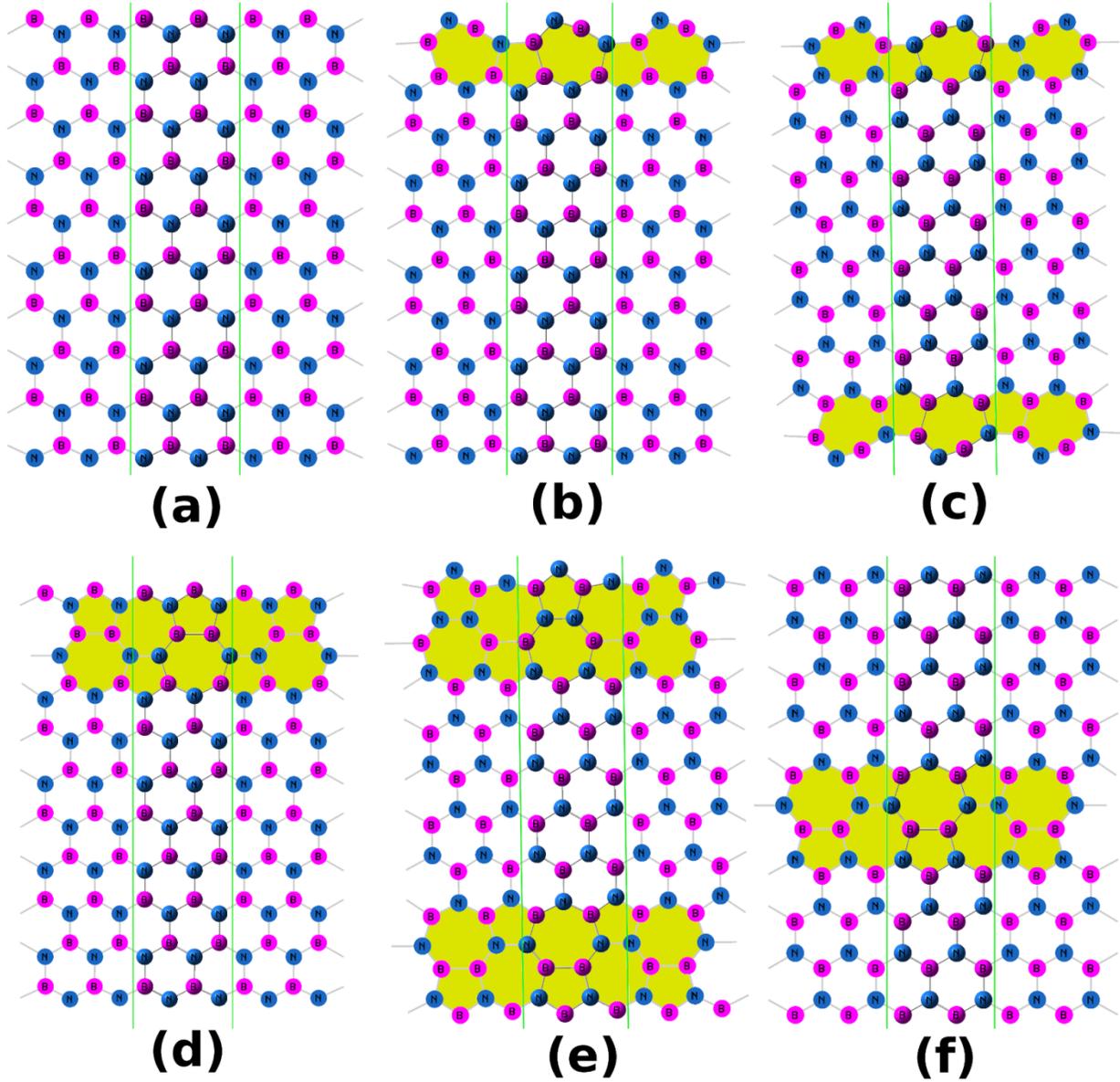

FIG. 1. (Color online) Structures of (a) perfect 10-zBNNR (b) De-ed-si-1_N (c) De-ed-bo-1 (d) De-ed-si-2_N (e) De-ed-bo-2 and (f) 4-D-3_De-mid-2. Shaded area represents the line-defect and the the unit cell of a system is the area inside the two staright lines (green color).

As the systems under consideration are quasi-one-dimensional nanoribbons, we have sampled the Brillouin-zone by $36 \times 1 \times 1$ k-points using the Monkhorst-Pack scheme[32], for the full-relaxation of the geometry, and $96 \times 1 \times 1$ k-points, for all the electronic and magnetic properties calculations. Systems are considered to be optimized only when the forces acting on all the atoms are less than 0.04 eV / Å. Cubic unit-cells with the initial lattice vectors (4.932, 0, 0); (0, 35, 0) and (0, 0, 15) Å, have been considered for all the systems of width ~ 2 nm (here, X-direction is considered as periodic and the width of the ribbon is considered along the Y-direction). After optimization, all the systems remained flat and a change in the lattice-vectors along the X-direction from 4.932 to ~ 5.01 Å has been found. For the width dependent studies, similar amount of vacuum (15 Å) has been considered in the non-periodic directions, as in the case of 2 nm ribbons, to avoid any spurious interactions between the nanoribbons and their periodic images. For all the Density of States (DOS) and projected-DOS (pDOS) plots, a broadening parameter of 0.05 eV has been used while plotting the energy levels.

III. SYSTEMS UNDER CONSIDERATION

Fig. 1 represents a few of various types of systems that we have considered in this study. We have named the systems with 5 indices. First index shows whether the ribbon is perfect (per) or it contains a defect (De). Second index shows whether the defect is at the edges (ed) of the ribbon or at the middle (mid) of the ribbon. Third index tells whether the defect is at single-edge (si) or at both the edges (bo). Fourth index tells the number of the PH-line-defect introduced. Fifth index specially represents those systems which have defect at one edge and are perfect at the other edge (see, for example, Figs. 1(b) and 1(d)) and this index distinguishes boron (B) and nitrogen (N) atoms at the perfect edge. The other two important types of systems (not shown in Fig. 1) are De-ed-si-1_B and De-ed-si-2_B, and they can be visualized by swapping the boron and nitrogen atoms of the systems in Figs. 1(b) and 1(d), respectively. It is required to mention, here, some *important and non-obvious points* regarding the PH-line-defects, which will be very much helpful to understand the results of the present work.

To generate an n-line-PH-defect, 'n' being an odd integer, we need 'n' number of zigzag and 'n' number of armchair chains (for example, 1-line-PH-defect can be visualized as a combination of 1-zigazag and 1-armchair chain. See shaded area in Fig. 1(b)). Importantly, any odd-line-PH-defect will have one edge as armchair-edge and the other edge will be zigzag (see the shaded area of Fig. 1(b)). In contrast to an odd-line-PH-defect, an even-line-defect always possesses the same type of edges (either armchair or zigzag) and generating an n-line-PH-defect, 'n' being an even integer, is divided into two cases: (i) If both-edges of the defect have to be armchair, then, we have to take (n-1)-zigzag and n-armchair chains and (ii) if both-edges of the defect have to be zigzag, then, we have to take n-zigzag and (n-1)-armchair chains (see the shaded area of Fig. 1(d) for an example of 2-line-PH-defect with both the edges being zigzag).

Notably, adding a single odd-line-defect at the edge of a nanoribbon will change the nanoribbon's edge nature, i.e. the edge will be changed from zigzag to armchair [compare Figs. 1(a) and 1(b)] and vice-versa. In contrast, introduction of a single even-line-PH-defect at the edge will not change the nanoribbon's edge nature [compare Figs. 1(a) and 1(d)]. Hence, we cannot (can) introduce a single odd-line-PH-defect (even-line-PH-defect) in a ribbon, by keeping both the edges of same kind. Finally, it should be noted that, any n-line-PH-defect can only be prepared from a ribbon with (n+1)-chains (for example, we can get at a maximum of a 9-line-PH-defect from a 10-zBNNR).

Keeping these facts in mind, we aimed at understanding how even and odd-line-PH-defects will affect the electronic and magnetic properties of a ribbon. For all the width independent studies, we have considered the zigzag-edged Boron-Nitride-nanoribbons of width ~ 2 nm, which according to the conventional notation[5] are equivalent to 10-zBNNRs (see Fig. 1(a)). In the 10-zBNNRs, we have varied the *number* of the PH-line-defect from one to eight. Also, as the *position* of a PH-line-defect can be either at an edge (single-edge or both-edge) or inside the ribbon, we considered the two-line-PH-defect as a representative system and we have varied its position in a 10-zBNNR from one edge of the ribbon to the other. Finally, to check how the effect of a defect would change with width, we have varied the width of the zBNNR from ~ 1.1 nm (6-zBBNR) to ~ 3.3 nm (16-zBNNR) with a 2-line-PH-defect. Spin-polarized DFT calculations, with different spin-configurations, have been performed for all these systems. Each spin-configuration is represented with an ordered pair, in which, the first element represents the spin-configuration at boron edge and the second element at the nitrogen edge. U and D represents the up and down-spins, respectively.

It is important to mention that, we didn't consider the systems, where 1-line-PH-defect is in the middle of the ribbon as a similar study has already been performed for GNRs[33] and also for BNNRs[34]. Also, systems with different types of defects at different edges (for example, 3-line at one-edge and 2-line at the other) are not considered. But, we strongly believe that one can easily expect the corresponding changes in the properties based on the present work.

## IV. RESULTS AND DISCUSSIONS

In this section, first we will compare the stability, spin-polarization and electronic and magnetic properties of the 10-zBNNRs with one and two-line-PH-defects (as representative systems for the odd and even-line-defects, respectively.) and also without any defect. Then, we will discuss the results of all other systems in a systematic manner.

### A. Perfect, 1-line-PH-defect and 2-line-PH-defect ribbons

In this work, we have calculated the formation-energy ($E_{Form}$) of the system as:

$$E_{Form} = E_{tot} - [nB * E_B + nN * E_N + nH * E_H] \qquad (1)$$

where, $E_{tot}$ is the total energy of the system; $E_B$, $E_N$ and $E_H$ are the total-energies per atom of α-boron, $N_2$-molecule and $H_2$-molecule, respectively. nB, nN and nH are the number of boron, nitrogen and hydrogen atoms in the system, respectively. $E_{Form}$ value of every system in every spin-configuration is given, with respect to the stable spin-configuration of that particular system and also with respect to the perfect 10-zBNNR [in (UD, UU) spin-configuration], respectively, in columns 3 and 4 of Table I. The negative formation energy for all the systems found in this work indicates that all are indeed stable.

From the Table I, one can find several interesting things. Firstly, in agreement with several previous studies on perfect zBNNRs, the spin-configuration with an anti-ferromagnetic ordering of the spins at the boron-edge and with a ferromagnetic ordering at the nitrogen edge, (UD, UU), was found to be the most stable spin-configuration. And, interestingly we find that, the same (UD, UU) spin-configuration is the most stable one for all the systems which have at least one zigzag edge with boron-atoms (i.e. irrespective of whether this zigzag edge arises because of a defect or not). Secondly, we find that the ground-state energy difference between the (UU, DD) and (UU, UU) spin-configurations, for all the systems, is always less than 5 meV and the energy difference between the (UD, UU) and (UU, DD)/ (UU, UU) is always large (in the range of 13 – 42 meV). The primary reason for these differences is the difference in the interaction between the spins of edge atoms. In the former case, although the edge spin-configuration has been changed from (UU, DD) to (UU, UU), the interaction between the spins on that edge didn't change (i.e. in both the cases the interaction is ferromagnetic), and hence, the change in energy is negligible. However, in the latter case, the spin-spin interaction at the edge has been changed from anti-ferromagnetic to ferromagnetic, and hence, the difference in the energy between the systems is appreciable. From above results, we can infer that, change in the system's energy will be largely dictated by the spin-spin interaction at the same-edge atoms, (rather than by the spin-spin interaction between different edge atoms) when the distance between the edges is large (here, ~ 2 nm), as shown previously for the case of z-BNNRs.[1, 3]

Thirdly, the stability of the systems decrease in the order of: De-ed-bo-1 > De-ed-si-1_N > De-ed-si-1_B > perfect-10-zBNNR >> De-ed-si-2_B > De-ed-si-2_N > 4-D-3_De-mid-2 >> De-ed-bo-2. This order can be explained based on the changes in the edge nature in these systems. As shown in Fig. 1, the system's edge nature changes from zigzag to armchair when we introduce a 1-line-PH-defect and it turns back to zigzag when we introduce a 2-line-PH-defect. Also, from the previous studies, it is known that the bare arm-chair edges are more stable than the bare zigzag-edges, in both zGNRs[24], and zBNNRs[35]. Combining these facts, we can see that, the higher stability attained by the ribbons after 1-line-PH-defect introduction is due to the defect's ability to bring an armchair-edge nature to a zigzag-edge. This is in agreement with the recent study by Bhowmick et. al., where they have shown that edge formation energy of the 5-7-edge-reconstructed z-BNNRs is 0.08 eV/Å less than the perfect z-BNNRs[10]. Similarly, as the 2-line-PH-defect didn't bring in any change to the edge nature, and also, as some energy has to be spent to reconstruct an edge, 2-line-PH-defect ribbons are less stable than perfect ribbons. The same reason holds for the stability order: De-ed-bo-1 > De-ed-si-1, and De-ed-si-2 > De-ed-bo-2. In the former case, there are two armchair edges and in the latter case, there are two zigzag edges, and hence, defect at both edges is more stable than the defect at a single edge. Also, the reason for 4-D-3_De-mid-2 to have similar energy as De-ed-si-2 is because both of them have a single-2-line-PH-defect. Next, we will consider the spin-polarization of these systems.

TABLE I. Spin-polarization ($S_{pol}$) of a system in different spin-configurations is given in column two. Columns 3 and 4 show the formation-energy of a system with respect to their corresponding stable spin-configuration and with respect to the perfect 10-zBNNR in (UD, UU) spin-configuration, respectively. Each spin-configuration is represented with an ordered pair, where, the first element of the ordered pair represents the spin-configuration at the boron-edge and the second element at the nitrogen edge. U and D represents up and down-spins, respectively. Energy of the most stable spin-configuration is scaled to zero.

| System name and Spin-configuration | $S_{pol}$ | Energy with respect to the most stable | Energy of the system with respect to the perfect 10-zBNNR in |
|---|---|---|---|

|  |  | conformer (meV) | (UD, UU) spin-configuration (eV) |
|---|---|---|---|
| Perfect 10-zBNNR, (UD, UU) (see Fig. 1(a)) | 1.926 | 0 | 0.000 |
| Perfect 10-zBNNR, (UU, DD) | 0.000 | + 16 | 0.016 |
| Perfect 10-zBNNR, (UU, UU) | 3.836 | + 16 | 0.016 |
| De-ed-si-1_B, (UD, UU) | 0.000 | 0 | -0.148 |
| De-ed-si-1_B, (UU, DD) | 2.000 | + 35 | -0.113 |
| De-ed-si-1_B, (UU, UU) | 2.000 | + 36 | -0.112 |
| De-ed-si-1_N, (UD, UU) | 0.000 | 0 | 0.283 |
| De-ed-si-1_N, (UU, DD) | 1.982 | + 21 | -0.262 |
| De-ed-si-1_N, (UU, UU) | 1.982 | + 21 | -0.262 |
| De-ed-bo-1, [(UD, UU), (UU, UU) and (UU, DD)] (see Fig. 1(c)) | 0.000 | 0 | -0.489 |
| De-ed-si-2_B, (UD, UU) | 1.921 | 0 | 3.911 |
| De-ed-si-2_B, (UU, DD) | 0.000 | + 13 | 3.924 |
| De-ed-si-2_B, (UU, UU) | 3.820 | + 13 | 3.924 |
| De-ed-si-2_N, (DU, UU) (see Fig. 1(d)) | 2.000 | 0 | 4.416 |
| De-ed-si-2_N, (DD, UU) | 0.000 | + 28 | 4.444 |
| De-ed-si-2_N, (UU, UU) | 3.764 | + 28 | 4.444 |
| De-ed-bo-2, (UD, UU) (see Fig. 1(e)) | 2.000 | 0 | 8.303 |
| De-ed-bo-2, (UU, DD) | 0.000 | + 39 | 8.341 |
| De-ed-bo-2, (UU, UU) | 3.745 | + 42 | 8.345 |
| 4-D-3_De-mid-2, (DU, UU) (see Fig. 1(f)) | 1.918 | 0 | 4.470 |
| 4-D-3_De-mid-2, (DD, UU) | 0.000 | + 22 | 4.492 |
| 4-D-3_De-mid-2, (UU, UU) | 3.818 | + 17 | 4.487 |

We have calculated the *Spin-polarization* of a system as:

$$S_{pol} = Q_{up} - Q_{down} \quad (2)$$

where, $Q_{up}$ ($Q_{down}$) is the Spin-up (spin-down) charge density. If $S_{pol}$ is non-zero (zero), the system is spin-polarized (spin-unpolarized) [with a spin magnetic moment, m = ($S_{pol}$) * $\mu_\beta$, where $\mu_\beta$ is the Bohr-magneton [36]]. Values of the $S_{pol}$ have been given in the Table I. Based on the results of our work, we propose that, the spin-polarization value of a BNNR for a particular spin-configuration can be easily understood, if we know the edge nature (i.e. zigzag or armchair) of the BNNR. In general, (1) if both the edges are zigzag, then it will have a finite spin-polarization, only (a) when both the edges have ferro-magnetic spin ordering and both the edges are coupled ferro-magnetically or (b) when one edge have a ferro-magnetic spin ordering and the other has anti-ferromagnetic spin-ordering. In the latter case, both the ferro-magnetic and anti-ferromagnetic coupling across the edges will lead to a finite spin-polarization. (2) If one edge is armchair and the other edge is zigzag, then the system will have finite spin-polariztion when the spins are ordered ferro-magnetically at the zigzag edge, irrespective of the spin-ordering at the arm-chair edge. Finally, (3) if both the edges are armchair, then the system will have zero-spin polarization, irrespective of the spin-ordering at the edges. For all the other cases, the systems will not be spin-polarized. All the above statements will be strictly valid for the bare BNNRs. Also, when the interaction between the two edges is negligible, the net spin-polarization of a system will be equivalent to the sum of the total amount of spin present at each edge. A similar type of conclusion has been drawn by Barone et. al. regarding the formation energy, where they have shown that when the edges are less interacting, then the formation energy of a particular configuration can be approximated as the summation of the formation-energies of the other individual spin-configurations[3]. Finally, it is important to mention that, the above statements are highly dependent on the passivation, and we expect that, the magnetic-moment which would arise from a bare zigzag edge will be completely lost when the edge is passivated with hydrogen. The reason for the observed $S_{pol}$ value for any system in Table I can be easily understood based on the above points.

Now, let us concentrate on the electronic and magnetic properties of these systems. Although the (UD, UU) is the most stable configuration for the bare z-BNNRs, a richer spectrum of electronic and magnetic behavior has been proved to be shown by the spin-configurations (UU, UU) and (UU, DD), both in the presence and absence of an external electric-field.[1,3] Following these studies, we have also plotted the projected-density-of-states (pDOS) for all the three different spin-configurations, as shown in Figs. 2 and 3. From both the Figs. 2 and 3, we find that, either a change in the spin-configuration or a change in the system leads to a change in the pDOS plots, primarily near the Fermi-level and these states near the Fermi-level have a major contribution from the edge-atoms of the nanoribbon. In the following, first we will describe the changes in the pDOS plots with a change in the spin-configuration for each system and then we will compare and contrast the pDOS plots of different systems.

*1. Perfect ribbons*

Fig. 2(a) shows the pDOS plots of the pristine 10-zBNNR in all the three different spin-configurations, which compares fairly well with the previous studies.[1,3] (The minor changes are mainly due to the difference in the widths of the z-BNNRs considered.[37]) From Fig. 2(a), we can notice that, a change in the spin-configuration of a nanoribbon is clearly reflected in the pDOS plot. We find that, when the spin-configuration on the nitrogen edge (N-ed) of the ribbon is changed from UU to DD (i.e. from (UU, UU) to (UU, DD)) the pDOS of the N-ed atoms changes drastically. Moreover, if we take a mirror image of the N-ed pDOS of the (UU, UU) configuration across the zero of the Y-axis, the result is the N-ed pDOS of the (UU, DD) configuration. This is because, the interaction between the spins at different edges is negligible, and hence, a change in the spin-configuration from UU to DD, at the N-ed, could only bring a change in the sign of the pDOS but not its magnitude. In contrast to the above, if we change the spin-configuration on a particular edge, we can expect a change also in the magnitude of the pDOS (this is because the distance between the spins on a single edge (~ 0.25 nm) is much less than the distance between the spins on two different edges (~ 2 nm)), and, this is what we have observed for the (UD, UU) spin-configuration of the perfect 10-zBNNR, as shown in the Fig. 2(a).

The major difference between the pDOS plots of (UU, UU) and (UD, UU) spin-configurations is the following (see Fig. 2(a)): B-ed pDOS which was spin-polarized and broad near the Fermi-level in (UU, UU) configuration has changed to non-spin-polarized and narrowed in (UD, UU) configuration. The change in the spin-polarization is because of the change in the spin-spin interaction between the edge-boron atoms [ferromagnetic (anti-ferromagnetic) in the former (latter) case]. The change in the width of the peaks of pDOS can be explained as follows: First, we should remember that the area under the B-ed pDOS for a particular energy range will give the number of B edge-states in that range and the number of edge-states of a system will be constant under any spin-configuration. Now, as we change the spin-configuration from (UU, DD) to (UD, UU) we bring in spin-symmetry at the B-edge, and this spin-symmetry urges equal contribution of the pDOS for both up and down spins. But, as the number of edge-states can't change with spin-configuration and as both-spins should have equal contribution, the broad peak has to narrow down without changing the area under the curve and this is what we have observed.

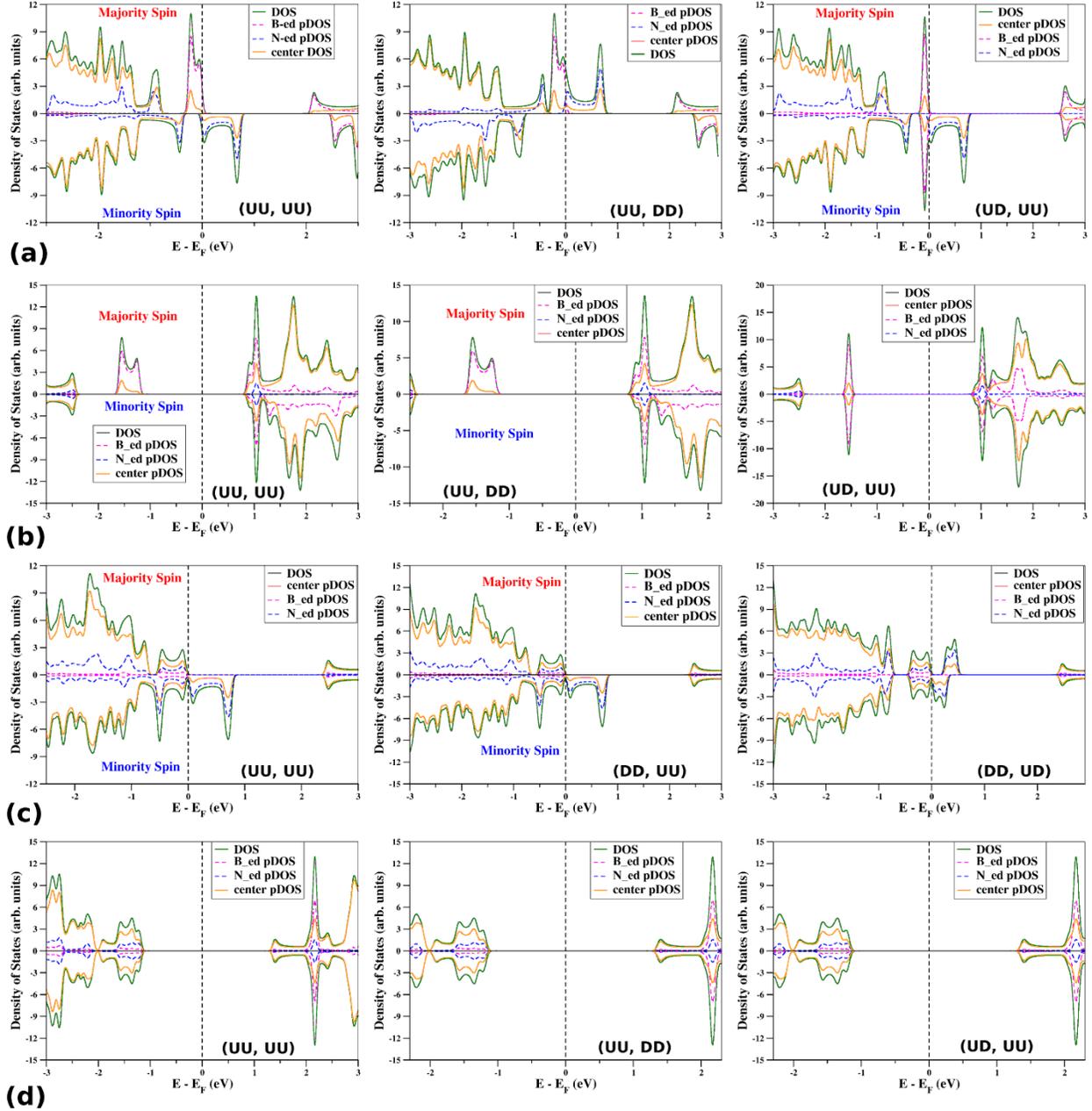

FIG. 2. (Color online) pDOS plots of each system in three different spin-configurations. (a) Perfect 10-zBNNR (b) De-ed-si-1_B (c) De-ed-si-1_N (d) De-ed-bo-1. Spin-configuration of a system has been labeled inside the pDOS plots. Majority and minority spins have been labeled whenever it is applicable. Solid lines with dark-gray (green) and light-gray (orange) colors show the complete DOS and the pDOS of all the atoms except the edge atoms, respectively. Dotted light-gray (magenta) and dark-gray (blue) color lines represent the pDOS of the edge boron and nitrogen atoms, respectively.

Finally, it is important to notice that, all the three magnetic-configurations are spin-polarized near the Fermi-level, and more importantly, among them the (UU, DD) shows half-metallic behavior, although its total spin-polarization is zero (see Table I and Fig. 2(a)). Thus,

**12**

10-zBNNR in (UU, DD) spin-configuration behaves as an *anti-ferromagnetic half-metal (AFHMs)*. Previously, in 1995, Groot et. al. have introduced this concept of AFHM[38] and later several compounds have been shown to possess this AF-half-metallicity as described by Wang et. al. [39] Among the several applications of an AFHM, its usage as a tip in the spin-polarized scanning-tunneling-microscope is very interesting. [38, 39] So, AFHM materials are surely useful in the future spintronics devices and to attain this AFHM property in a system without containing any element with d-orbitals, i.e. in 10-zBNNR, is very much interesting. Lastly, it is surprising to see that, existing studies [1, 3] on zBNNRs with (UU, DD) spin-configuration haven't discussed the AFHM behavior, even though their results indicate it!

*2. One-line-PH-defect ribbons*

In addition to the previous sub-section, where we have shown the effect of spin-configuration, in this sub-section, we will show how the introduction of a line-defect can bring in changes to the electronic and magnetic properties of a system. Figs. 2(b) and 2(c) show the pDOS plots of the systems De-ed-si-1_B and De-ed-si-1_N, respectively. From the pDOS plots, it is clear that systems have transformed from spin-polarized metals/half-metals to spin-polarized semi-conductors (for (UU, UU) and (UU, DD) configurations) after the introduction of a one-line-PH-defect at one-edge of a perfect ribbon. The spin-polarization in these systems is solely due to the spin alignment at the perfect edge. The zero-contribution of the defect-edge towards the spin-polarization of the system can be clearly understood by noticing the pDOS of the edge-atoms. For example, from Fig. 2(b), one can notice that, both the valence band maximum (VBM) and the conduction band minima (CBM) have a major contribution from the edge boron states which are present at the perfect edge. The fact that the spin-polarization is only due to the perfect-edge can be proved by comparing the pDOS plots of (UU, UU) and (UU, DD) configurations, where, we have changed the spin-configuration on a defect-edge from UU to DD and we couldn't find any change in the pDOS plots. Whereas, when we changed the spin-configuration on the perfect-edge from UU to UD (now, compare (UU, UU) and (UD, UU) configurations), we can see that the system has lost its spin-polarization (which is because of the anti-ferromagnetic coupling between the spins at the perfect (boron)-edge). Finally, when one-line-PH-defect is introduced at the remaining perfect-edge (i.e. De-ed-bo-1), the system transforms from a spin-polarized semi-conductor to non-magnetic insulator. This result compares quite well with the results of Bhowmick et. al.,[10] where they have shown that 5-7 reconstruction at both the edges of a zBNNR is detrimental to the magnetic moment of the ribbon. From the above discussion, we can notice that, one can tune the system properties from metallic to semiconducting to insulating, by introducing a 1-line-PH-defect at the edge of a zBNNR.

*3. Two-line-PH-defect ribbons*

Although, we have proved, in the previous sub-section, that the introduction of a 1-line-PH-defect at a zigzag-edge destroys the magnetic moment, we didn't show whether this behavior is universal for any PH-line-defect or not. Indeed, in this sub-section we will prove that, a z-BNNR will lose (hold) its magnetic moment if the introduced line-defect has both armchair and zigzag-edges (only zigzag-edges), like 1-line-PH-defect (2-line-PH-defect). Figs. 1(d)–1(e) show some example systems whose edge nature retains even after the introduction of a PH-line-defect.

Fig. 3 shows the pDOS plots of De-ed-si-2_B, De-ed-si-2_N, De-ed-bo-2 and 4-D-3-De-mid-2 systems. All these systems have finite spin-polarization near the Fermi-level and all of them are metallic at least for one spin, the other varied depending on the system and spin-configuration. So, unlike 1-line-PH-defects, 2-line-PH-defects will preserve the magnetic nature of the systems. Similar to the perfect ribbons (see Fig. 2(a)), these systems also change their edge-pDOS near the Fermi-level depending on their spin configuration and the changes are quite similar to the perfect ribbons. For example, in Fig. 3a we find a mirror image behavior of the N-ed pDOS between the spin-configurations, (UU, UU) and (UU, DD), which is exactly what we have seen in Fig. 2(a). Similarly, if we compare the (UU, UU) and (UD, UU) configurations of Fig. 3(a), we can find the loss in spin-polarization for the B-ed pDOS in (UD, UU) spin-configuration because of the expected anti-ferromagnetic coupling between the boron-edge atoms. A similar behavior in the pDOS plots has been found for the other systems (see Fig. 3).

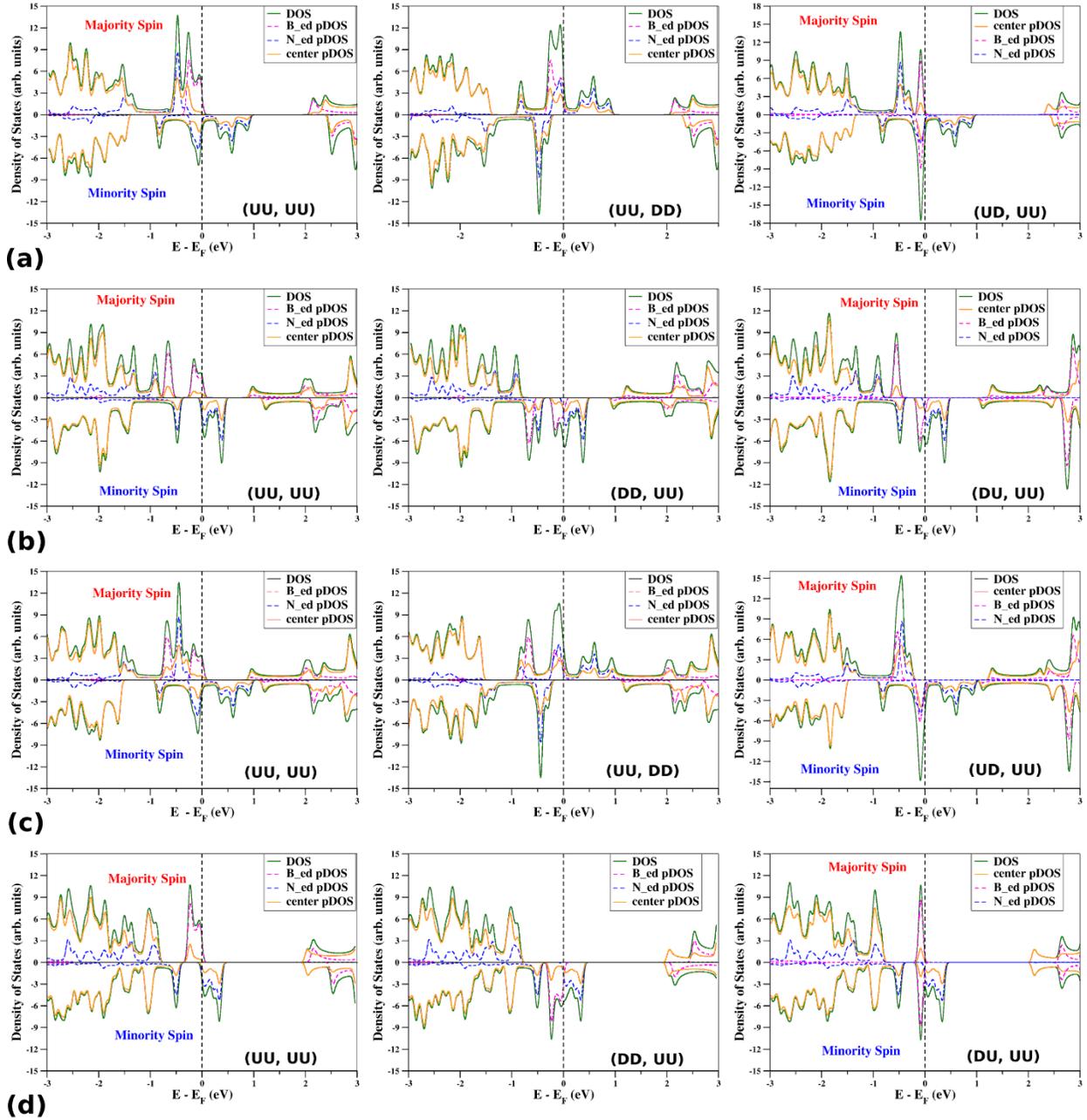

FIG. 3. (Color online) pDOS plots of each system in three different spin-configurations. (a) De-ed-si-2_B (b) De-ed-si-2_N (c) De-ed-bo-2 (d) 4-D-3-De-mid-2. Spin-configuration of a system has been labeled inside the pDOS plots. Majority and minority spins have been labeled whenever it is applicable.

An important finding which might not be very much obvious on the first look at Fig. 3 is: "Although the B-ed pDOS is non-spin-polarized in the (UD, UU) [or (DU, UU)] configuration for the systems, De-ed-si-2_B and 4-D-3-De-mid-2 [here after we call set-A], it remain spin-polarized for the De-ed-si-2_N and De-ed-bo-2 systems [here after we call set-B]". The main
**15**

reason for the above finding is that boron atoms in the zigzag-edges of the systems in set-A and set-B are different from each other. Systems in set-A have a zigzag-edge which is made up of only hexagons i.e. a perfect-edge (see Fig. 1(f)), whereas, systems in set-B have a zigzag edge which is formed by alternating fused heptagons and pentagons i.e. a defect-edge (see Figs. 1(d) and 1(e)). So, in the former case, both edge-boron atoms belong to hexagon rings of the zigzag-edge, whereas in the latter case, one edge-boron atom belongs to a pentagon-ring and the other belongs to a heptagon-ring. This structural difference in the edge-boron atoms between the two sets is the reason for the difference in the spin-polarization found here. To prove this further, we have plotted the pDOS for each boron edge-atom for the systems De-ed-si-2_N and De-ed-si-2_B in (DD, UU)/ (UU, DD) configuration, as representative candidates for the defect and perfect edge systems, in Figs. 4(a) and 4(b), respectively.

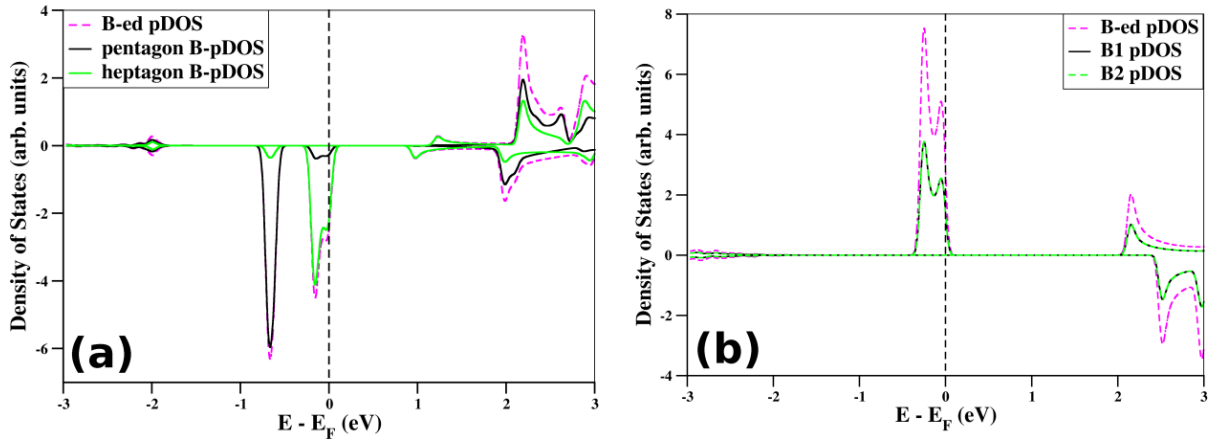

FIG. 4. (Color online) pDOS plots of each boron edge atom for the systems (a) De-ed-si-2_N and (b) De-ed-si-2_B in (UU, DD) spin-configuration. Pentagon [heptagon] B-pDOS means, pDOS of the edge boron atom which belong to the pentagon [heptagon] ring. B1, B2-pDOS indicates the pDOS from two different hexagon edge-boron atoms, which we numbered as 1 and 2.

Fig. 4(b) shows the individual pDOS contributions of each edge-boron atom to the total B-ed pDOS for De-ed-si-2_B system (a system which has boron atoms at the perfect-edge) and from the figure, we can notice that all the B-ed pDOS peaks have equal contribution from both the edge atoms. Firstly, as all the peaks have a contribution from each edge atom, a change in the spin on any edge-atom will have an effect on all the peaks. This is the reason to get a change in the peak width when we change the spin on the boron-edge atoms from UU to UD. Next, as the contribution to each peak from each edge atom is same, peaks will have a zero-spin-polarization (finite spin-polarization) if spin on one edge atom is different (same) from the spin on the other

**16**

atom. This is the reason to find a zero-spin-polarization (finite spin-polarization) for B-ed pDOS in the UD (UU) configuration. Contrary to the perfect edge systems, each peak in the B-ed pDOS (see Fig. 4(a)) of the defect edge-systems (here, De-ed-si-2_N) has a major contribution either from a pentagon or from a heptagon edge-boron atom (but not the same contribution from both the edge atoms). Now, as all the peaks have contributions from both the edge atoms (although differ in magnitude), here also, all peaks will change with a change in spin on an edge atom. This explains the reason for the changes in the peak widths for a change in spin-configuration from UU to UD. Next, as the contribution from both the edge-atoms is not same for any peak, each peak will have a finite spin-polarization, irrespective of the spin on the each edge-boron atom. This is the reason for the spin-polarization found even after a change in the spin-configuration on the B-ed atoms from UU to UD in the defect edge system. This is also the reason for the non-equivalent changes in the peak widths for the up and down spins of B-ed pDOS.

From all the above discussions, we can conclude that (i) System's spin-configuration can change its electronic and magnetic properties (ii) Among the several possible spin-configurations, perfect zBNNRs in (UU, DD) spin-configuration can act as potential candidates for anti-ferromagnetic half-metals, and hence, might be useful in the preparation of spin-polarized STM tips. Also, after comparing Figs. 2 and 3 we find that, whenever a system's edges are zigzag (both edges) then the system shows half-metallicity, only in the (UU, DD) spin-configuration. (iii) Introduction of a 1-line-PH-defect at an edge would destroy that edge's magnetic moment, and it is due to the armchair-edge generated by the 1-line-PH-defect, (iv) unlike the 1-line-PH-defect, a 2-line-PH-defect will retain the edge magnetism and spin-polarization of a zBNNR due to its zigzag edge nature. But, unlike perfect zigzag-edge, a zigzag-edge formed by the introduction of a 2-line-defect's will be comprised of a pentagon and heptagon. And, due to this, the changes in the electronic and magnetic properties of a defective zigzag-edge system, with a change in spin-configuration, are quite different from that of a perfect-zigzag edge system. Finally, (v) by knowing the pDOS of one spin-configuration, we can easily guess the pDOS of any other spin-configuration, provided (a) the spin-spin coupling between the atoms at two different edges of the ribbon is negligible and (b) we know the edge nature of the system (i. e. perfect edge or defect edge).

To find whether the above conclusions regarding the 1-line and 2-line-PH-defects are true for all cases, we have varied the number of line-defect from 3 to 8 in a 10-zBNNR. Here after, we will present the results only for the (UU, DD) spin-configuration as this configuration shows promising properties. The results are presented below.

TABLE II. Spin-polarization ($S_{pol}$) of each system is given in column two. Column 3 shows the stability of a system with respect to the perfect 10-zBNNR in (UU, DD) spin-configuration and is represented as $E_{form}$.

| System-name | $S_{pol}$ | $E_{Form}$ (eV) |
|---|---|---|
| Perfect-10-zBNNR | 0.000 | 0.000 |
| **Odd-line-PH-defects** | | |
| De-ed-si-1_B | 2.000 | -0.129 |
| De-ed-si-1_N | 2.000 | -0.278 |
| De-ed-si-3_B | 2.000 | 4.261 |
| De-ed-si-3_N | 1.988 | 4.607 |
| De-ed-si-5_B | 2.000 | 9.350 |
| De-ed-si-5_N | 1.978 | 8.664 |
| De-ed-si-7_B | 2.000 | 13.426 |
| De-ed-si-7_N | 1.992 | 13.193 |
| **Even-line-PH-defects** | | |
| De-ed-si-2_B | 0.000 | 3.908 |
| De-ed-si-2_N | 0.000 | 4.428 |
| De-ed-si-4_B | 0.000 | 8.292 |
| De-ed-si-4_N | 0.000 | 8.836 |
| De-ed-si-6_B | 0.000 | 12.627 |
| De-ed-si-6_N | 0.000 | 13.174 |
| De-ed-si-8_B | 0.000 | 16.933 |
| De-ed-si-8_N | 0.000 | 17.408 |
| **Variation in the defect position** | | |
| 0-D-7_De-mid-2 | 0.000 | 3.908 |
| 1-D-6_De-mid-2 | 0.000 | 4.411 |
| 2-D-5_De-mid-2 | 0.000 | 4.476 |
| 3-D-4_De-mid-2 | 0.000 | 4.473 |
| 4-D-3_De-mid-2 | 0.000 | 4.476 |
| 5-D-2_De-mid-2 | 0.000 | 4.478 |
| 6-D-1_De-mid-2 | 0.000 | 4.467 |
| 7-D-0_De-mid-2 | 0.000 | 4.428 |

B. Odd and even-line-defect ribbons

Table II shows the spin-polarization of the 10-zBNNRs with different line-defects in (UU, DD) spin-configuration. Clearly, all the odd-line-defects are spin-polarized and all the even-line-defects are zero-spin-polarized. As explained earlier (for 1 and 2-line-defect ribbons in section A), the reason for the finite spin-polarization for all the odd-line-defect ribbons is the presence of a single zigzag-edge which holds the spin, and the reason for zero total spin-polarization of the even-line-defects is because of the presence of two such zigzag edges, which carry exactly the same amount of spin but with opposite sign. Thus we prove that, the $S_{pol}$ of the system is mainly due to the zigzag-edge of the system (irrespective of whether it is a perfect or a defect edge).

Table II also shows the stability of all these systems [with respect to the perfect-10-zBNNR in its (UU, DD) spin-configuration]. $E_{Form}$ values of all these systems prove that, (i) not only the nature of the defect (i.e. odd or even-line) but also the size of the defect (i.e. 4-line or 8-line) will change the system stability and (ii) *for a ribbon with constant width, greater the number of the line-defect, lesser is its stability*. We have also verified our statement regarding the gain in stability due to the presence of an armchair-edge in 1-line-PH-defect ribbon, by comparing the $E_{Form}$ values of every odd-line-defect ribbon with its immediate lower even-line-defect ribbon [i.e. for example, we have compared the $E_{Form}$ of 7-line-defect ribbon with the $E_{Form}$ of the 6-line-defect ribbon] and we find that the $E_{Form}$ values of these systems are nearly equal (with in the order of an eV). (Note that, comparing a 7-line-defect with 8-line-defect will not give any new understanding on the armchair-edge nature, because, 7-line-defect will obviously be more stable than 8-line-defect for its lesser size. On the other hand, 6-line-defect is expected to be more stable than 7-line-defect just based on its lesser size. So, a comparable $E_{Form}$ value of 7-line-defect with a 6-line-defect indicates the armchair-edge's ability to stabilize the system). But, the amount of the stability acquired by a system through a change in the edge nature from zigzag to armchair might not be equal to the required amount of energy for the edge-reconstruction. If the former energy dominates, then the system attains stability and if latter energy dominates, the system will be unstable. The domination of the latter energy is the reason for the lesser stability of higher odd-line-defect (i.e. from 3-line) systems compared to the 1-line-PH-defect systems, even though they possess an armchair edge.

Along with the stability and spin-polarization, we have also calculated the changes in the band-structure with a change in the defect line-number (n). Figs. 5 and 6 show the band-structure of all the odd and even-line-defects in (UU, DD) spin-configuration, respectively. From the band-structure plots, we find that all the systems with boron atoms at perfect-edge behave in one-

**19**

kind and with nitrogen atoms at perfect-edge behave in another kind. This finding is valid for both even and odd-line-defect systems. For example, near the Fermi-level, all the odd-line-defect systems with boron atoms at the perfect-edge have nearly similar behaviors in dispersion, band-gap etc. (see Fig. 5). Similar behavior can also be seen for even-line-defects (see Fig. 6). These findings prove that, "it is the edge nature of the defect which mainly dictates the electronic properties of the system than the number of the defect-line (n)".

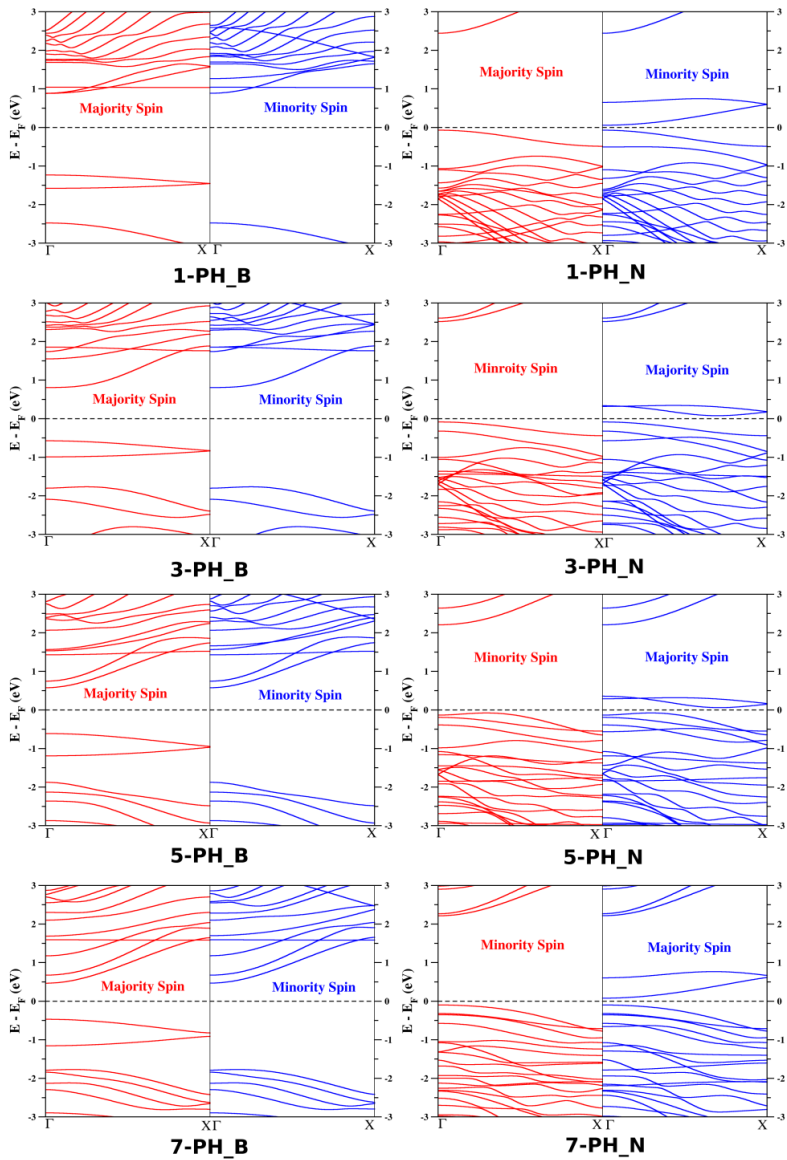

FIG. 5. (Color online) Band-structure plots of all odd-line-defect ribbons. The number of the line-defect (n) and atoms at the perfect edge (either B or N) are indicated below each plot, as n-PH_B/N. In each plot

**20**

up-spin (down-spin) is given on the left-side (right-side). Majority and minority spins have been labeled whenever it is applicable.

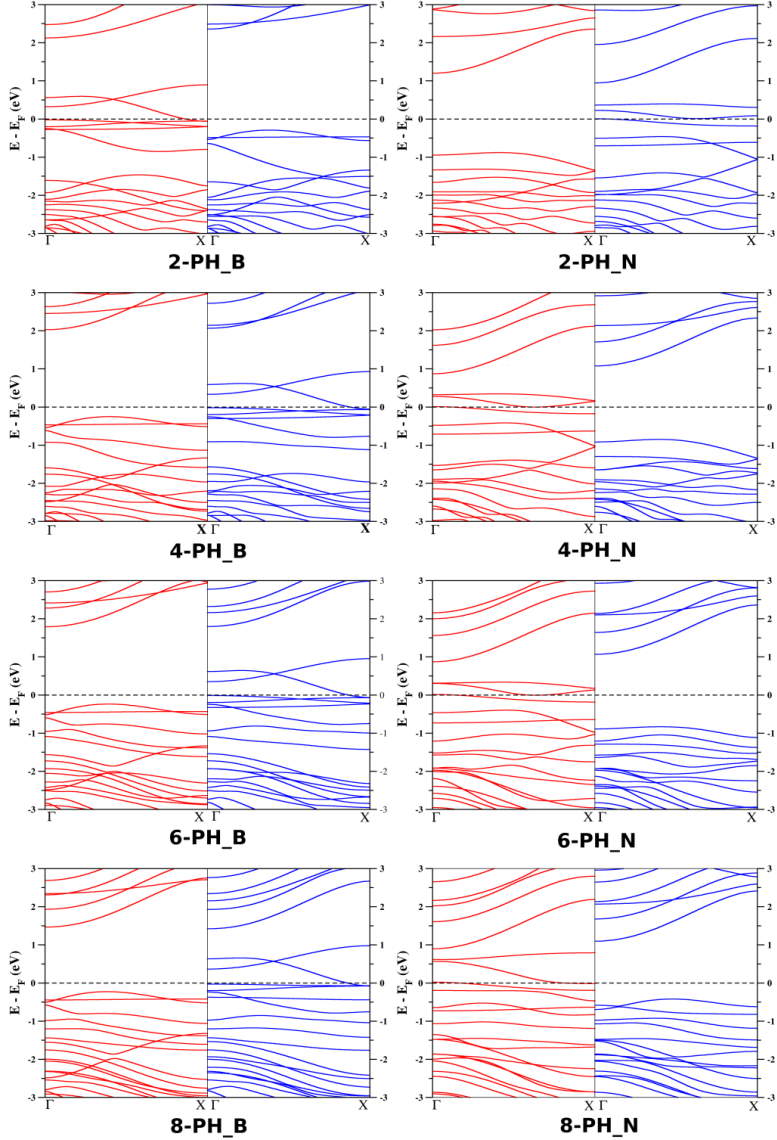

FIG. 6. (Color online) Band-structure plots of all even-line-defect ribbons. The number of the line-defect (n) and atoms at the perfect edge (either B or N) are indicated below each plot, as n-PH_B/N.

C. Variation of the defect-position

In this sub-section, we will discuss how the variation in the position of a defect affects the system's property. For this, we have chosen 2-line-PH-defect as our representative candidate and

**21**

we have varied its position. We have chosen "2-line-PH-defect" because: 1) 2-line-PH-defect is an even-line-defect, and hence, irrespective of its position in the ribbon, it will not change the edge nature of the system, and also 2) according to our studies, 10-zBNNR with 2-line-PH-defect shows half-metallicity and we want to know whether this half-metallicity is retained even after a change of the defect position. As we already know that, a 2-line-defect will have 3 zigzag chains and as the ribbon considered has 10-zigzag chains, we are left with only 7-perfect zigzag chains after the introduction of a 2-line-PH-defect. These 7 chains can be arranged across the defect in 8 ways, and hence, there are 8 systems to study. The nomenclature of these systems is n-D-m_De-mid-2, where n, m represents the number of zigzag chains the defect (D) is away from the nitrogen and boron edge, respectively. For example, 2-D-5_De-mid-2 means, the defect is 2-chains away from the nitrogen edge and 5-chains away from the boron-edge.

$E_{Form}$ values of all the eight systems are given in the Table II and except for the case of 0-D-7_De-mid-si-2-pr *there is hardly any change in the formation energy with change in the defect position*. This shows that, unlike the defect nature (i.e. odd-line or even-line), defect position has less effect on the formation energy, unless it changes the edge nature. The reason for the "0-D-7" system to be different from others is because of its nitrogen edge. In 0-D-7, the zigzag-edge with nitrogen atoms has the defect, whereas, for all the other systems, the nitrogen edge is perfectly zigzag-edged. One interesting point to notice here is that, while translating the defect from B-ed to N-ed, the maximum energy was consumed only when the edge has changed from perfect zigzag-N-edge to defect zigzag-N-edge (i.e. from 1-D-6_De-mid-si-2-pr to 0-D-7_De-mid-si-2-pr), which again proves the importance of the edge nature in stabilizing a structure. As both the edges are zigzag and as the spin-configuration is (UU, DD), all these systems show zero $S_{pol}$. Fig. 7 shows the band-structure plots of all the eight systems. Clearly, all the systems are half-metallic irrespective of the defect position, which again proves that, the reason for the half-metallicity is the zigzag-edge nature of these ribbons (which is preserved in all of them even though the defect is moving from one-end to the other-end of the ribbon).

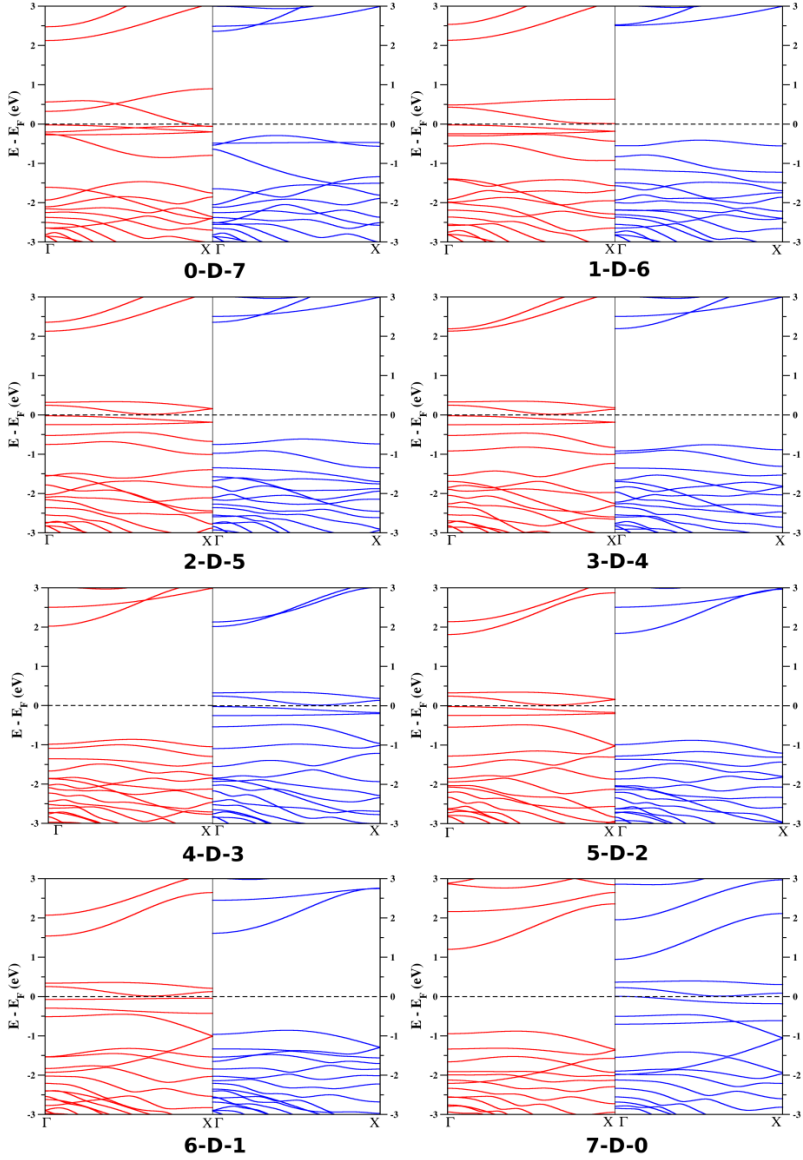

FIG. 7. (Color online) Band-structure plots of 10-zBNNR with a 2-line-PH-defect at different positions inside the ribbon. The position of the defect is indicated using the index n-D-m (see the text for nomenclature).

D. Varying the width of the ribbon:

Finally, we have varied the width of the ribbon from 6-zBNNR to 16-zBNNR with a 2-line-PH-defect to check the robustness of our results. Again, $S_{pol}$ value for all the systems is zero (and hence, not shown) because of their zigzag edge nature. Table III shows the formation energy for all the widths, which we have considered. As the number of atoms in each system varies, we have given the formation-energy values per atom, ($E_{Form}$), for each one of these

**23**

systems. These values show that, *formation of a defect is energetically more favorable in a bigger ribbon than in a smaller ribbon*. Importantly, the change in the $E_{Form}$ energy decreases and reaches almost a *saturation value* (less than room-temperature), once the system has more than 15-chains (> 3.1 nm). For example, the difference in the $E_{Form}$ values between the ribbons with '6 and 7' or '7 and 8' chains is ~ 0.1 eV, whereas, between '14 and 15' or '15 and 16' chains is just 0.02 eV.

TABLE III. Variation in the formation energy, $E_{Form}$ (eV/ atom), with a variation in the number of zigzag chains, n, of De-ed-si-2_N systems is given.

| n | 6 | 7 | 8 | 9 | 10 | 11 | 12 | 13 | 14 | 15 | 16 |
|---|---|---|---|---|----|----|----|----|----|----|----|
| $E_{Form}$ | -0.962 | -1.061 | -1.136 | -1.195 | -1.241 | -1.280 | -1.311 | -1.338 | -1.362 | -1.382 | -1.399 |

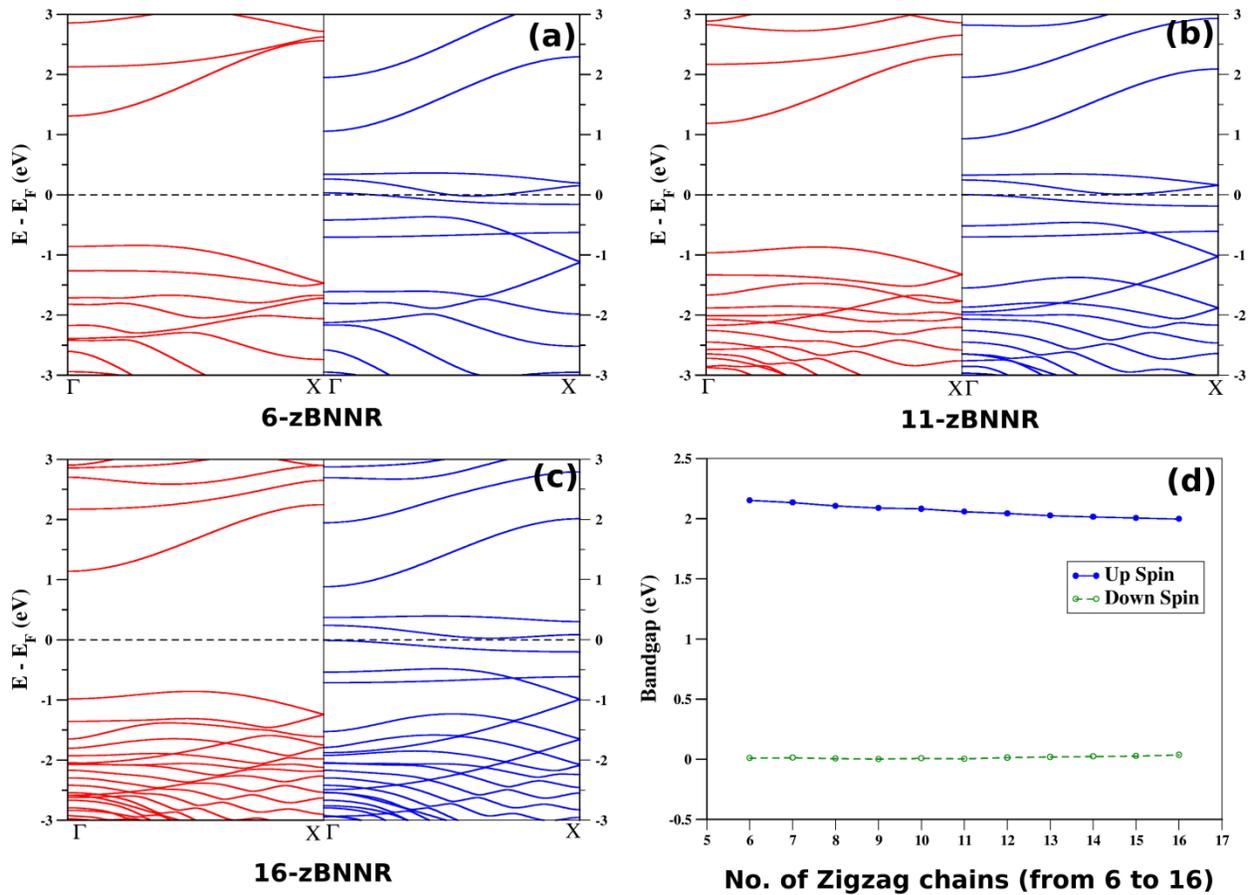

FIG. 8. Band-structure plots of (a) 6-zBNNR (b) 11-zBNNR and (c) 16-zBNNR. (d) Variation of the band-gap with width, for both up and down-spins.

In Figs. 8(a)–8(c) we have shown the band-structure plots for three representative widths, viz., 6, 11 and 16-zBNNRs, respectively. Note that, the band-structure plots for all other widths considered in this study show exactly the similar behavior. From Figs. 8(a)–8(c), we find that the half-metallicity in 10-zBNNR is in fact robust against the ribbon width. These plots also show that, the dispersive nature of the states near the Fermi-level hasn't changed much with the ribbon width. Finally, in Fig. 8(d), we have presented the band-gap variation with a change in the ribbon width. As shown, with an increase in the width of the ribbon, there is very small change in the band-gap of the system's for both up and down-spins. Thus, this plot also proves the robustness of half-metallicity in 2-line-PH-defect systems.

## V. CONCLUSIONS

In conclusion, we have shown that all the properties of the systems presented in this work, are dependent mainly on the edge-nature of the ribbon, and we have also shown, the edge nature of a ribbon can be tuned using PH-line-defect number. Among the several spin-configurations which we have considered, we find (UD, UU) configuration to be the most stable one, and the (UU, DD) configuration to be interesting one because it displays anti-ferromagnetic half-metallic behavior. We also find that, in contrast to the presence of a pair of hexagons at the perfect-zigzag-edge, a defect-zigzag-edge consisting of a pentagon and heptagon pair can cause the large differences in their DOS, when the spin-configuration at these edges are changed.

Within the (UU, DD) spin-configuration, we have shown that, a system with, (i) one-edge as zigzag and other edge as arm-chair (odd-line-defects) will behave as spin-polarized semi-conductors, and (ii) both edges being zigzag (perfect ribbons or even-line-defect ribbons) will behave as anti-ferromagnetic half-metals. We have proved the robustness of the half-metallicity of the zigzag-edged systems against the defect line number, position of the defect and width of the ribbon. We have also discussed the stability of the ribbons and we have shown that, introduction of an n-line-defect is energetically more favorable for smaller 'n' than for larger 'n' and for a particular 'n', larger the size of the ribbon, it is energetically easier to introduce the defect. Finally, we conjecture that, half-metallic nature which we find for (UU, DD) spin-configuration is quite robust and would be observed for all the systems (with/without impurities, defects etc.) which have zigzag-edge-nature at both the edges.

ACKNOWLEDGEMENTS

SKP acknowledges DST (Govt. of India) and AOARD, US Air Force for research grants.

REFERENCES

\* Corresponding author: pati@jncasr.ac.in